\documentclass[11pt]{emulateapj}
\usepackage{natbib}
\usepackage{graphicx}% Include figure files

\newcommand{\be}{\begin{equation}}
\newcommand{\ee}{\end{equation}}
\newcommand{\ba}{\begin{eqnarray}}
\newcommand{\ea}{\end{eqnarray}}
\def\lesssim{\mathrel{\hbox{\rlap{\hbox{\lower4pt\hbox{$\sim$}}}\hbox{$<$}}}}
\def\grtsim{\mathrel{\hbox{\rlap{\hbox{\lower4pt\hbox{$\sim$}}}\hbox{$>$}}}}

\begin{document}
\title{A detection of dark matter halo ellipticity using galaxy cluster lensing in SDSS}
\author{Anna Kathinka Dalland Evans\altaffilmark{1}, Sarah
  Bridle\altaffilmark{2}}
\altaffiltext{1}{
  Institute of Theoretical Astrophysics, University of Oslo, Box 1029, 0315 Oslo, Norway.
}
\altaffiltext{2}{
Department of Physics and Astronomy, University College London, Gower Street, London, WC1E 6BT, UK.
}
\begin{abstract}
  We measure the ellipticity of isolated clusters of galaxies in the
  Sloan Digital Sky Survey (SDSS) using gravitational lensing. We
  stack the clusters, rotating so that the major axes 
%Krev of the cluster members 
of the ellipses determined by the positions of cluster member galaxies %Krev
are aligned. We exclude the signal from the central 0.5
  $h^{-1}$ Mpc to avoid problems with stacking alignment and cluster
  member contamination. We fit an elliptical NFW profile and find 
%Krev an axis ratio 
a projected, two-dimensional axis ratio %Krev
for the dark matter of $f = b/a = 0.48^{+0.14}_{-0.09} \
  (1 \sigma)$, and rule out $f=1$ at 99.6 per cent confidence thus
  ruling out a spherical halo. We find that the ellipticity of the
  cluster galaxy distribution is consistent with being equal to the
  dark matter ellipticity. The results are similar if we change the
  isolation criterion by 50 per cent in either direction.
\end{abstract}


\section{Introduction}
Cosmological simulations can be used to predict many different
statistics of the mass distribution in the Universe. The most commonly
employed statistic is the two-point correlation function, or its
Fourier counterpart the power spectrum. Three-point statistics are
much harder to predict and measure, and higher orders are rarely
discussed. A more popular statistic is the number of peaks in the mass
distribution, as given by the number of clusters of galaxies. The dark
matter power spectrum and the number of clusters of galaxies are often
cited as among the best ways to constrain the properties of dark
energy~\citep[e.g.][]{detf}.

Uncertainties on cosmological parameters are decreased when two
measurements have different parameter degeneracies and are often
referred to as `complementary'. %Krev
In this paper we consider a statistic which %Krev should offer
may offer %Krev
complementary constraints on cosmology: the \emph{shapes} of peaks in the
mass distribution, as probed by the ellipticity of galaxy cluster dark
matter halos. In addition this may place important constraints on
modifications to the law of gravity since we may compare the results
from both dark and light matter, as also tested by studying the dark
and light matter distributions in the bullet
cluster~\citep{clowebgmrjz06}.

Predictions of cluster ellipticities come mostly from numerical simulations
\citep{west89,
detheije95,
jing02,
floor03,
ho04,
flores05,
rahman06}.
The ellipticity
is expected to depend on
cosmological parameters
\citep{evrard93,
splinter97,
buote97,
suwa03,
rahman04}
and to evolve with redshift \citep{kasun05, allgood06},
an evolution which itself might depend on cosmology
\citep{hopkins05,
ho06}.
The distribution of sub-halos within a cluster halo is found to be an
indicator of the overall halo ellipticity \citep{bode07} but is
slightly less elliptical.

The ellipticity of the brightest cluster galaxy and the ellipticity of
the distribution of cluster member galaxies is much easier to observe
than any other cluster ellipticity measure. This has been studied by a
number of authors
\citep{west90,
plionis91,
rhee91,
strazzullo05}.
Most recently, \cite{wang07} studied groups of galaxies in SDSS and
found an alignment between the brightest cluster galaxy
(BCG)
and the distribution of galaxies which was strongest in the most
massive groups and
between red BCGs and red group member galaxies.

In this paper we focus on measuring the dark matter ellipticity
directly using gravitational lensing. We also compare this ellipticity
to the ellipticity of the cluster member galaxy distribution, to see
how reliable a tracer of ellipticity the light is, and therefore
whether it can be used by itself for cosmological studies.

Gravitational lensing has been used very successfully to measure the
mass and profile of clusters of galaxies by many authors. The work
most relevant to our study is that of \cite{sheldon01,sheldonI07} and
\cite{sheldonIII07} who stack the lensing signal from many clusters of
galaxies to find an average signal. \cite{nat2000} proposed that when
stacking the shear signal from many halos the stack could be made
while retaining information about the major axis of the halo, as
observed from the distribution of light. The stacked shear map should
then provide a constraint on the ellipticity of the dark matter halo,
if indeed the mass and light were aligned. They considered
lensing by galaxies but we apply the same technique to clusters of
galaxies here.

The ellipticity of the dark matter distribution has been studied
observationally for galaxy sized halos using gravitational lensing 
using data from the Red-Sequence Cluster Survey %Krev
\citep{hoekstrayg04},
%Krev and
the Sloan Digital Sky Survey
\citep{mandelbaumea06}
and the Canada-France-Hawaii Telescope Legacy Survey \citep{parker07}. %Krev
This has proved to be extremely difficult and we present here, for the
first time, results from stacking galaxy cluster halos.
\cite{cypriano04}
has previously
made measurements of the cluster dark matter ellipticity using
gravitational lensing measurements from individual clusters. They
found a good agreement between the dark matter halo orientation and
the orientation of the brightest cluster galaxy.

When calculating angular diameter distances and the mean density of
the Universe we assume a flat cosmology with $\Omega_m = 0.3$.  The
NFW halo profile has a very weak dependence on the fluctuation
amplitude. We assume $\sigma_8=0.8$ for this calculation.

The structure of this paper is as follows. In Sec.~\ref{sec:data} we
describe our data set and operations we have performed on the data
such as rotating and stacking and removal of neighboring clusters.  In
Sec.~\ref{sec:modelling} we use two theoretical models - a Singular
Isothermal Ellipsoid (SIE) and a Navarro-Frenk-White \citep[NFW, see
][]{nfw} model - of the mass and light distribution, and look at
correction factors from the redshift distribution and cluster
decontamination. Our results, and some interpretations of these, are
presented in Sec.~\ref{sec:results}.  Conclusions are summarized and
discussed in Sec.~\ref{sec:conclusions}.

\section{Data}
\label{sec:data}

In this section we describe the catalogues used and the operations we
carried out before comparing with models, including cluster selection
and stacking and rotating.

\subsection{Catalogues}
We use the cluster catalogue of \cite{bencat}, which is at the present
time the largest existing galaxy cluster catalogue, consisting of
13,823 galaxy clusters from the Sloan Digital Sky Survey
\citep{york00}. The cluster galaxies are red-sequence members
(occupying the so-called E/S0 ridgeline in colour-magnitude space),
brighter than $0.4 L_*$ in the $i$ band and between redshifts $0.1 < z
< 0.3$.  
%The cluster redshifts are photometric, with uncertainties of
%$\sim 0.01$ \citep{bencat}. %Krev
They also lie within a circular aperture of radius
$R_{200}$, which is the estimated radius within which the density 
%Krev is 200 times that of 
%Krev the critical density. 
of galaxies with $24 < M_r < 16$ is
200 times the mean density of such galaxies. %Krev
The number of cluster members
inside this aperture is $10 \le N_{\rm{gals}}^{r200} \le 188 $; the
lower limit is a requirement for inclusion into the catalogue. We
refer to \cite{bencat} for details of the catalogue and to
\cite{benmaxbcg} for a description of the cluster selection algorithm.

In order to define an isolated sample, we removed clusters found to be
too close to each other, as seen on the sky. Details on this close
neighbour removal can be found in Sec.~\ref{sec:closen}.  Clusters
that are too close to the survey edge are also removed, with the
requirement that the minimum distance from the cluster center to the
survey edge is $7 h^{-1}$ Mpc.  We are left with a total of 4281
clusters to analyse for our purposes.  We have also organised the
clusters into 4 redshift bins between $z=0.10$ and $z=0.30$, each of
width 0.05 in $z$. 
% based on the photometric redshifts, using a photometric redshift width of
%0.05. 
Since these cluster redshifts are photometric, with uncertainties
of $\sim 0.01$ (Koester et al 2007a), the spectroscopic bin width
will be very slightly larger than the photometric bin width. We do not
take into account this small broadening in our analysis because we
expect it to have negligible effect on the ellipticity results.%Krev
%Krev We have also organised the clusters into 4 redshift bins between
%Krev $z=0.10$ and $z=0.30$, each of width $0.05$ in $z$.

The shear galaxy catalogue is the same as used in \cite{sheldon04},
except that the area covered is larger, $\sim 6325 \ {\rm
  deg}^2$. There is approximately one galaxy per square arcminute in
this catalogue. Galaxies in the shear catalogue have
extinction-corrected $r$-band Petrosian magnitudes less than 22. Stars
have been removed from the catalogue by the Bayesian method discussed
in \cite{scranton02}. Unresolved galaxies and objects with photo-$z$
errors greater than $0.4$ have also been removed. To correct the
shapes of galaxies for effects of PSF dilution and anisotropy, the
techniques of \cite{bernstein02} were used, with modifications
specified in \cite{hirata03}. We refer to \cite{sheldon04} for full
details of the compilation of the shear catalogue.%Krev

We do not want to include in the shear catalogue galaxies which are
already in the cluster member catalogue. Therefore we remove from the
shear catalogue all galaxies which are close, as seen on the sky, to a
cluster member. We cut on a physical (as opposed to angular) distance
of 0.012 $h^{-1}$ Mpc, as calculated at the redshift of the cluster.
This distance corresponds to 5 arcseconds at redshift $z = 0.2$. In
principle this cut may also remove some real background galaxies,
however the shear measurements of these galaxies will in any case
likely be adversely affected by the light contamination by the cluster
members. Note that we do not eliminate \emph{all} cluster members from
the shear catalogue with this cut, only those clusters that fit the
selection criteria set by the maxBCG selection method 
%(red galaxies,brighter than 0.4 L*; Koester et al. 2007a).
\citep[red galaxies, brighter than 0.4 L*;][]{bencat}.

\vspace{0.5cm}
\subsection{Postage stamp size and close neighbor removal}
\label{sec:closen}

In order to avoid contamination of the shear signal from neighboring
structures, we only include in our analysis galaxies which are
sufficiently close to a cluster, as seen on the sky. We decide this
distance by considering the predicted contribution of neighboring
clusters to the shear signal, and then use this to decide the required
separation of neighboring clusters.

We include in our shear catalogue galaxies in a square postage stamp
of $10 \times 10 \ h^{-1}$ Mpc centered on each cluster. This choice
was made using Fig.~8 of \cite{johnston}, which shows model fits for
the lensing signal split up into contributions from several
components. We want the lensing signal in our postage stamp to be
dominated by the central cluster, not by the contribution from
neighboring mass concentrations such as nearby clusters and filaments.
Our interest therefore lies in comparing their NFW profile (green
line) to the contribution from neighboring halos (blue line).  The
results in their Fig.~8 are shown for 12 richness
($N_{\rm{gals}}^{r200}$) bins. The mean of the $N_{\rm{gals}}^{r200}$
values of our cluster sample is $ \sim 24$, directing us to the panel
$N_{\rm{gals}}^{r200} [18-25]$. We match our postage stamp limit at
the radius where the NFW contribution is roughly equal to that of the
contribution from neighboring clusters. Note that we are also removing
close neighbors (see below), so the contribution from neighboring
halos would actually be even lower for our case, so our cut may be
conservative.

Clusters that are \emph{too} close to each other in projected
separation represent a challenge for our analysis. The mass
distribution of a cluster will affect the shear field of its neighbor.
This could have a significant effect on the measured ellipticity.
While it is possible to isolate clusters in three dimensions within a
simulation, we are plagued by overlaps on the sky since gravitational
lensing measures the projected mass. We therefore endeavor to remove
this complication by selecting relatively isolated clusters for this
analysis. For proper comparison with theory, a similar sample should
be made from simulations. However, in this paper we are primarily
concerned with the first significant detection of ellipticity in a
large sample.

We do not want the shear from a neighbor to appear within our postage
stamp, but this is unavoidable to some extent, because the shear at
one point is affected by mass far away (shear is non-local). The
effect on the measured ellipticity of having neighbors is twofold: (i)
It can make the distribution more circular, if the neighbor position
is uncorrelated with cluster major axis. This will occur due to chance
alignments close to the line of sight, particularly from clusters in
different redshift bins; (ii) It can make the distribution more
elliptical, if the neighbor position is correlated with major axis
direction. Pairs of clusters which are physically close will probably
be aligned along the major axis of the cluster
\citep[e.g.][]{plionis91}.

We therefore remove clusters that are closer than $5 h^{-1}$ Mpc
(corresponding to half the width of the postage stamp) to neighboring
clusters.  When a cluster has one or more neighbors with an angular
separation corresponding to less than $5 h^{-1}$ Mpc, calculated at
the middle of the redshift bin of the cluster (see later), we discard
it if any of the neighbors have a higher $N_{\rm{gals}}^{r200}$. The
neighbor removal was done consecutively from low to high redshift.
This process reduced the number of available clusters to 4281. In a
later section, we analyze the use of different minimum close neighbor
distances.

Although we remove a cluster's less rich neighbors from our sample,
the shear pattern of the remaining cluster will already have been
affected by its neighbor(s). However, by retaining the richest of the
neighboring clusters we hope that the shear field is dominated by
this cluster.

%\clearpage
\begin{figure*}
\scalebox{.283}{\includegraphics[width=24cm]{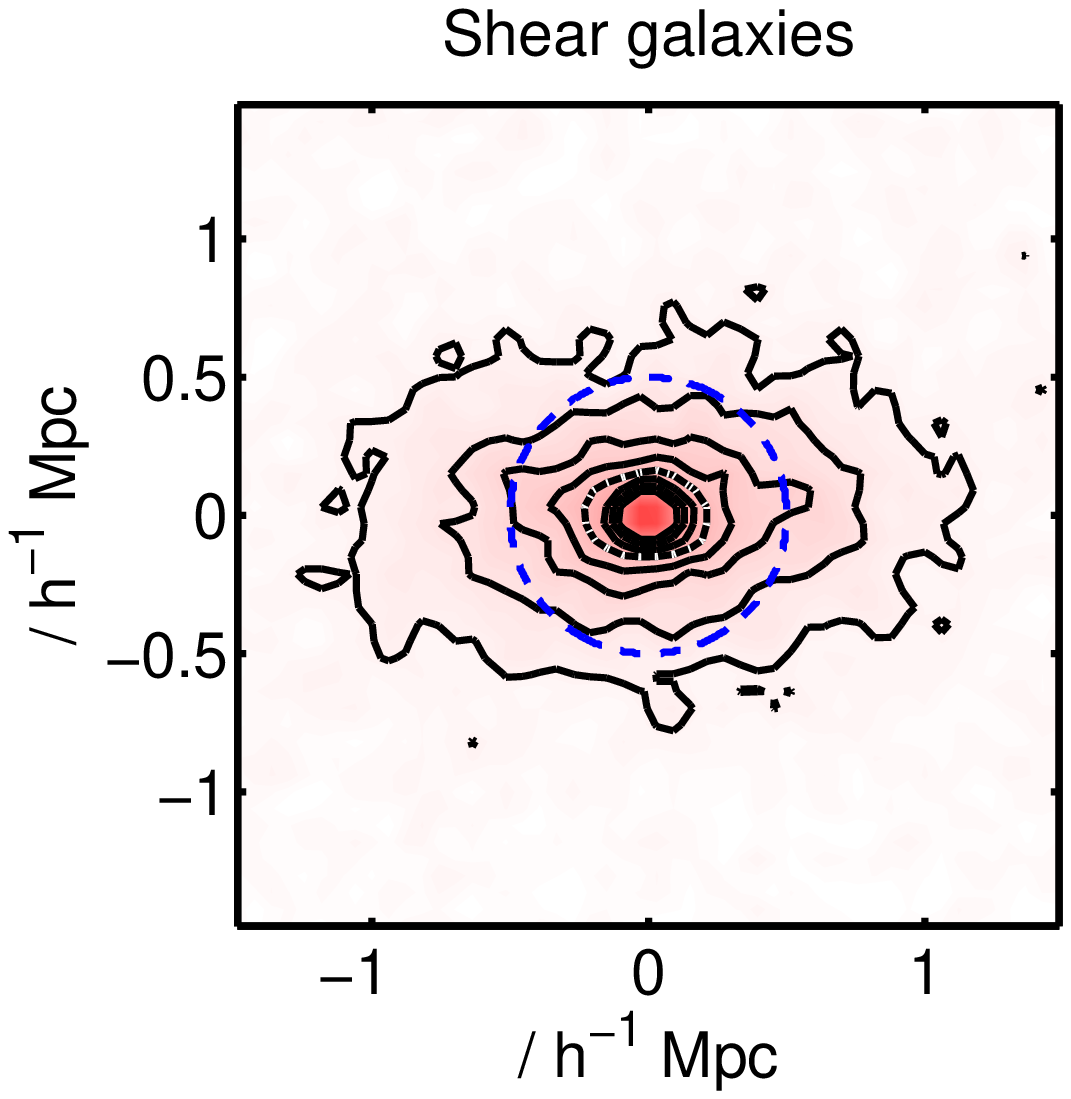}}
\scalebox{.255}{\includegraphics[width=24cm]{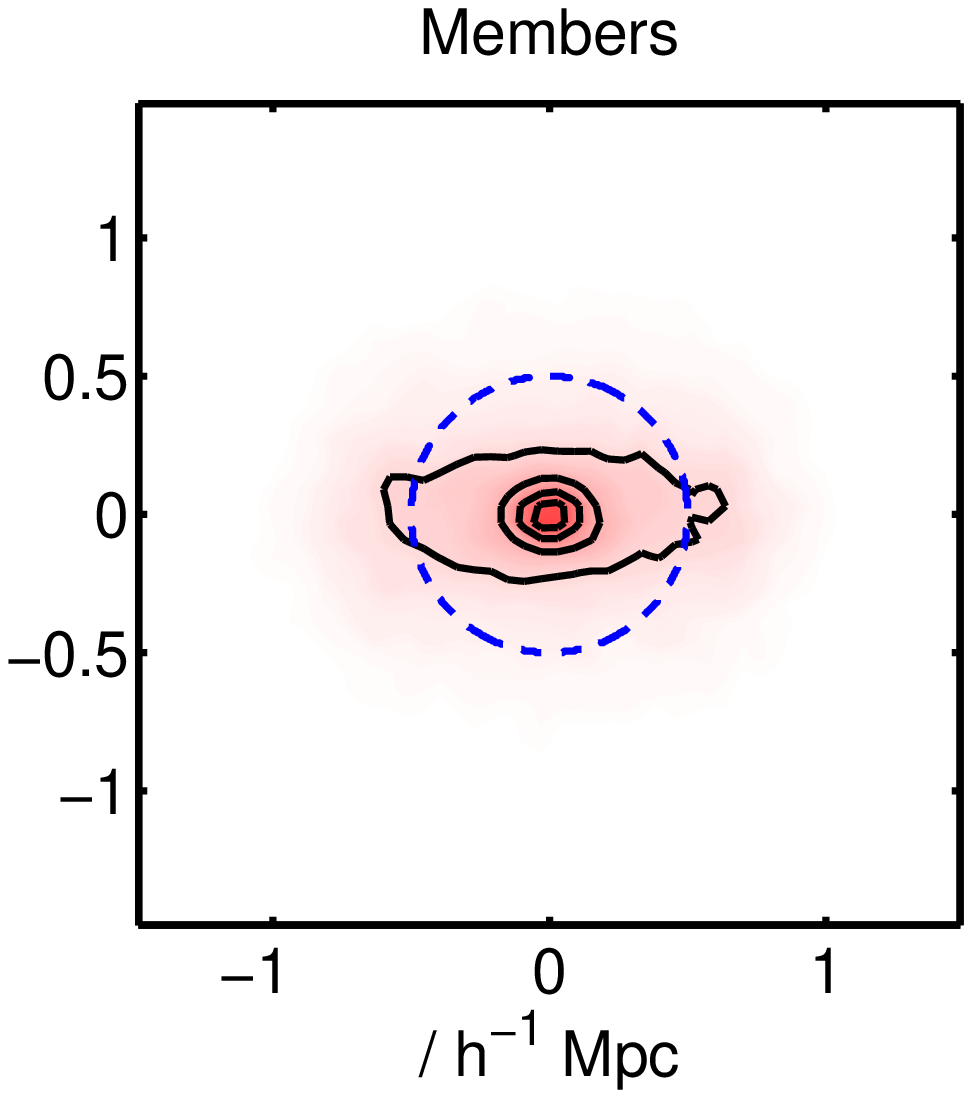}}
\scalebox{.255}{\includegraphics[width=24cm]{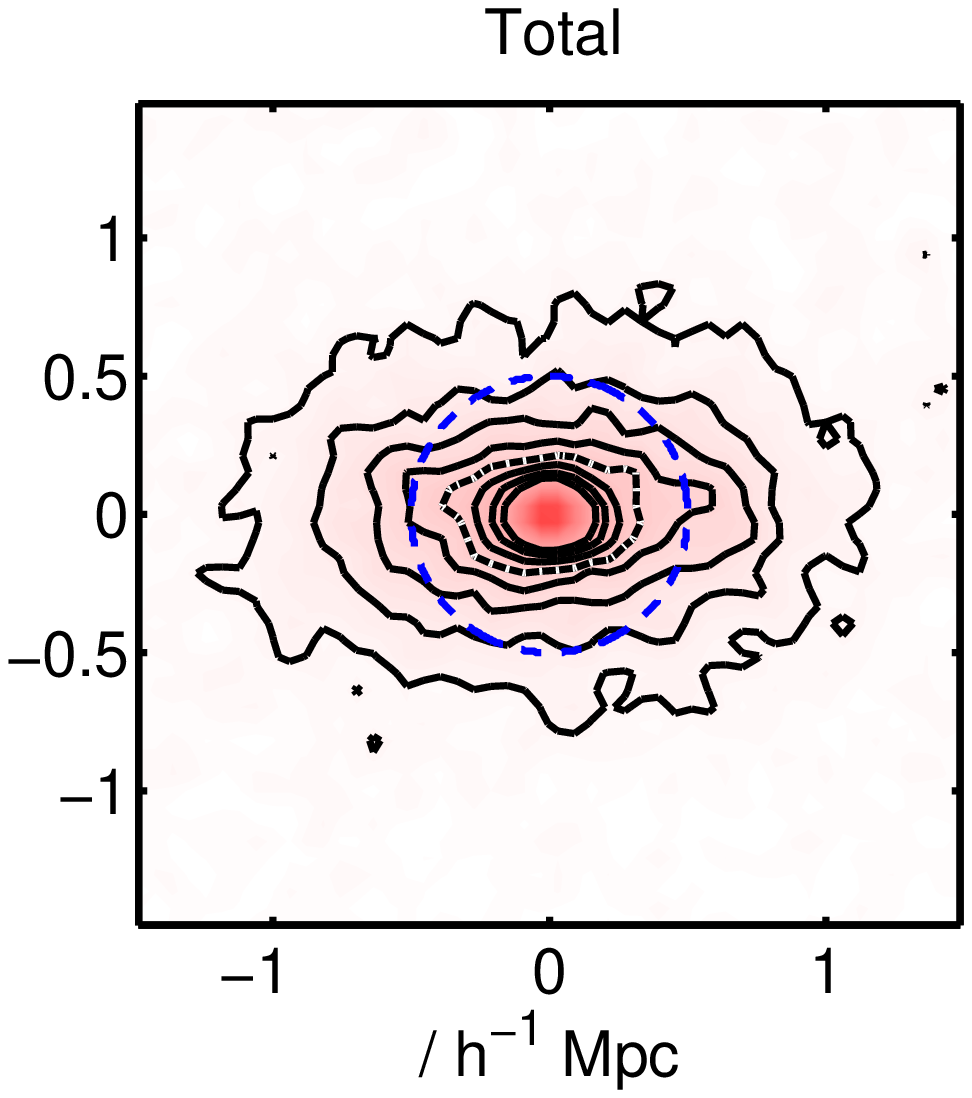}}
\caption{ Total number of galaxies per unit area for redshift bin 3
  ($0.20 < z < 0.25$) for illustration. From left to right: shear
  galaxies, cluster members and total galaxies (shear galaxies plus
  cluster members). The left and right panels include galaxies which
  are uncorrelated with the cluster.  We draw our contours relative to
  the constant background level (which is the same number in both left
  and right panels, since the cluster member catalogue does not
  contribute).  For the left and right panels, contours are equally
  spaced from 1.2 to 2.6 times the background level, in steps of 0.2.
  The dashed contour shows 2 times the background level.  For the
  central panel the contours are 0.2, 0.4, 0.6, 0.8 times the same
  background level.  The dashed circle shows location of mask, see
  text. For comparison, a $3\times10^{14} h^{-1} M_{\odot}$ cluster at
  a redshift of 0.2 has $R_{200} \sim 1.4 h^{-1}$ Mpc, where $R_{200}$
  is the radius within which the mean density of the cluster is 200
  times the mean matter density at the cluster redshift. We have
  zoomed in to show the central part of the postage stamp for clarity.
}
  \label{fig:numbgals}
\end{figure*}
%\clearpage

\subsection{Stacking and rotating}
\label{sec:stackandrot}

On average, we have only around 1 shear galaxy per square arcminute
%Krevand each galaxy estimates the shear with an %Krev accuracy 
%Krevuncertainty %Krev
and the uncertainty on the shear for a single galaxy is an order of magnitude
larger than the shear we are
trying to measure.  Therefore we need to use the shear signal from
many clusters in order to obtain a significant signal. We therefore
stack the clusters on top of each other to improve the signal-to-noise
ratio. In other words, we use information from the postage stamp field
of shear galaxies for all clusters simultaneously.

The stacking could be carried out in either physical or angular space.
For our method the two approaches are exactly equivalent if the
redshift bins are small enough. We stack in angular space and use
redshift bins of redshift width 0.05. This causes a radial blurring
because two clusters of the same physical size will be stacked on top
of each other in angular space to have different angular sizes.  The
blurring of the shear and light maps will be of at worst plus and
minus 20 per cent (for the lowest redshift bin).  Since superposing
ellipses of different scalings retains the original ellipticity, this
results in a slightly smoother cluster profile but will not affect our
ellipticity results. We find that our results are fairly similar even
when comparing two very different profiles, see
Sec.~\ref{sec:modelling}.

Straightforward stacking of elliptical clusters with random
orientations would erase any ellipticity and produce a circular
average cluster. Before stacking, we therefore rotate each cluster to
lie along an $x$-axis, which is the major axis as defined by the
ellipticity of the cluster members, see Fig.~\ref{fig:numbgals}. This
rotate-and-stack method is described by \cite{nat2000} for use in
galaxy-galaxy lensing, and we have, for the first time, applied this
technique for use on cluster lensing.

We calculate the direction of a cluster major axis from the positions
of the cluster members, as defined in the cluster catalogue. We do not
take into account the luminosity of each cluster member. The cluster
center $(x_c,y_c)$ was taken to be the position of the brightest
cluster galaxy (BCG) as defined by the maxBCG algorithm
\citep{benmaxbcg}. However, the position of the BCG is not necessarily
coincident with the cluster's actual centre of mass. For comparison,
we therefore calculate the center of each cluster as given by the mean
position of the cluster members. The mean physical offset (for
redshift bin $0.20 < z < 0.25$) between the two center definitions is
$\sim 0.15 h^{-1}$ Mpc, and the standard deviation $0.09 h^{-1}$ Mpc.
Compared to our mask radius of $0.5 h^{-1}$ Mpc, therefore, the shift
in centre position is relatively small. Any effect this mis-centering
does have will increase the ellipticity of the members, which we do
not focus on here, and cause the misalignment angle to tend towards
the direction from the BCG to the center of the cluster member
distribution. This would in itself be an alternative and potentially
useful way to stack the clusters to obtain the results presented
here. Therefore we do not consider this effect further.

To find the ellipticity angle of rotation of the cluster, we use the
quadrupole moments of the cluster members.
The quadrupole moments are given by:
\begin{eqnarray}
Q_{\rm xx} & = & \langle (x_i - x_{\rm c})^2 \rangle_i \\
Q_{\rm xy} & = & \langle (x_i - x_{\rm c}) \  (y_i - y_{\rm c}) \rangle_i\\
Q_{\rm yy} & = & \langle (y_i - y_{\rm c})^2 \rangle_i
\end{eqnarray}
where the summation $i$ is over the cluster members. We convert this
into the ellipticity components $e_1$ and $e_2$ of the cluster through
the relations:
\begin{eqnarray}
\label{eq:e1}
e_1 & = & \frac{Q_{\rm xx} - Q_{\rm yy}} {Q_{\rm xx} + Q_{\rm yy} + 2
  \sqrt{Q_{\rm xx} Q_{\rm yy} - Q_{\rm xy}^2} }\\
e_2 & = & \frac{2 Q_{\rm xy}} {Q_{\rm xx} + Q_{\rm yy} + 2 \sqrt{Q_{\rm xx} Q_{\rm yy} - Q_{\rm xy}^2} } \,.
\label{eq:e2}
\end{eqnarray}
The angle of the cluster, anticlockwise from positive $x$ axis, is
then:
\be
\label{eq:thetarot}
\theta^{\rm{rot}} = \frac{1}{2}\ {\rm atan} \Bigl(
\frac{e_2}{e_1} \Bigr) \, .
\ee

We rotate the positions of the cluster member galaxies and the
positions of the shear galaxies using the following transformation
\begin{eqnarray}
x^{\rm{rot}} & = & d \ \rm{cos} ( \theta - \theta^{\rm{rot}} )\\
y^{\rm{rot}} & = & d \ \rm{sin} ( \theta - \theta^{\rm{rot}} )
\end{eqnarray}
where $(d,\theta)$ are the polar coordinates of the galaxy to be
rotated, relative to the cluster center. For clusters with
ellipticities close to zero, $\theta^{\rm{rot}}$ in
Eq.~\ref{eq:thetarot} has little meaning. From the cluster members
alone, 18\% of our clusters have an ellipticity ($e = \sqrt{e_1^2 +
  e_2^2}$, using Eqs.~\ref{eq:e1} and \ref{eq:e2}) of less than 0.1
and only 5\% have ellipticities less than 0.05. Therefore the angle is
reasonably well defined. 2\% of the clusters have an ellipticity
greater than 0.5. The cluster selection criteria by \cite{bencat} and/or
our isolation criteria have therefore done a reasonable job of
identifying isolated clusters.  In addition the clusters seem
relatively undisturbed, i.e.~have low ellipticity.  This also
illustrates that it would not be particularly useful to bin the
clusters according to the cluster member ellipticity because the range
is relatively small (and our final signal to noise is quite low).%Krev

%\clearpage
\begin{figure}
\includegraphics[width=8cm]{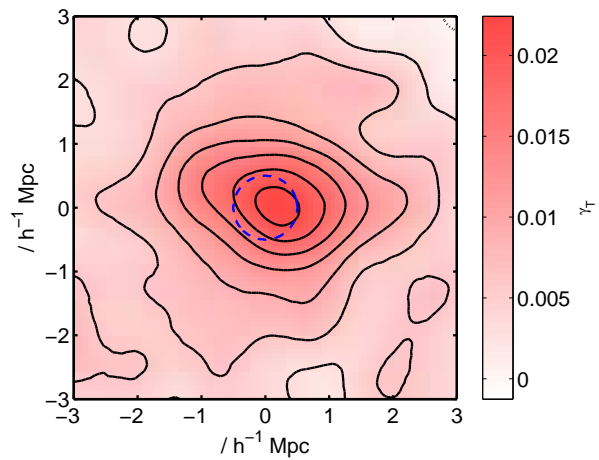}
\caption{
  Tangential shear,
  $\hat{\gamma}_{\rm{T}}$,
  as measured
  from all
  galaxies in the background galaxy catalogue
  %Krevfor the entire postage stamp of $10 h^{-1}$ Mpc.
  for a zoomed-in view of the postage stamp. %Krev
  To reduce the noise we have smoothed with a Gaussian with standard
  deviation $1 h^{-1}$ Mpc.  The location of the mask is shown by the
  dashed circle.  
}
\label{fig:gammaT}
\end{figure}
%\clearpage

Fig.~\ref{fig:numbgals} shows the number of galaxies per unit area,
for z bin 3, after stacking and rotating. The left panel shows
galaxies from the shear galaxy catalogue, with cluster member
catalogue galaxies removed, the middle panel shows cluster galaxies
from the cluster member catalogue and the right panel shows the sum of
shear galaxies and cluster members.

The left hand panel clearly contains a significant number of cluster
members, despite the fact that the galaxies from the cluster member
catalogue are not included. As described in \cite{benmaxbcg}, cluster
members for this catalogue were identified using the maxBCG algorithm.
This algorithm employs the red sequence method, based on the
observational fact that cluster galaxies occupy a narrow region (a
so-called \emph{ridgeline}) in a colour-magnitude diagram. This method
is designed to conservatively select red galaxies at the central area
of a cluster. As an illustration we investigate the central region of
the stacked cluster in the third redshift bin ($0.2<z<0.25$). In the
central square arcminute the number of galaxies in the member
catalogue is $ \sim 40 \%$ of the number of galaxies in the (members
removed) shear catalogue, after subtracting the constant background
level i.e. most of the cluster members are not in the cluster member
catalogue.

The existence of these extra members allows a very convenient check on
our stacking and rotating: the rotation angles were calculated from
the cluster member catalogue alone, whereas the left hand panel does
not include the galaxies used to decide the rotation angle. Therefore
the fact that we see ellipticity in this panel means that the angle
calculated from the members catalogue is correlated with the angle of
the extra cluster members. If, for example, we had calculated rotation
angles from a very small number of galaxies from the cluster member
catalogue, there would be a large degree of randomness due to shot
noise, and the alignment of measured and true ellipticity would be
random to a large extent, resulting in a circular pattern in the left
panel of Fig.~\ref{fig:numbgals}.

The object of this paper is to calculate the ellipticity of the dark
matter, as measured from gravitational lensing, and compare it with
that of the light-emitting galaxies. Any misalignment in the stacking
and rotating will tend to make the dark matter appear less elliptical.
However this misalignment will have the same effect on the ellipticity
of the light, as measured from the extra cluster members (left hand
panel) alone. Therefore we can compare like with like, and assess the
\emph{relative} ellipticity of the dark and light matter, despite any
misalignment.

We have chosen the contour levels so that a fair comparison can be
made between the left and central panels: the outermost contour
corresponds to the same cluster member density in each. Therefore we
see that at large radii most of the cluster members are not included
in the cluster member catalogue.  For all the axis ratio measurements
reported in this paper we exclude the central regions (see
Sec.~\ref{sec:contam}), this is shown by the dashed circle in
Fig~\ref{fig:numbgals}.  Outside this excluded region we see that the
contours change only a little from the left to the right panel. This
is convenient because it means that it is not too important whether we
compare the dark matter ellipticity with the light ellipticity derived
from either the left or the right hand panel.

We calculate the tangential, $\hat{\gamma}_T$, and cross,
$\hat{\gamma}_X$, components of the shear $\hat{\gamma}$ for each
shear galaxy
\begin{eqnarray}
\hat{\gamma}_{\rm T}    & = & \hat{\gamma} \ {\rm cos} (2 \alpha)  \\
\hat{\gamma}_{\rm X} & = & \hat{\gamma} \ {\rm sin} (2 \alpha)
\end{eqnarray}
where $\alpha$ is the angle between the shear galaxy major axis and a
tangent through the center of the shear galaxy, with respect to the
cluster center.  We calculate the shear estimate for each galaxy from
the polarizations ($\epsilon$) %Krev
in the shear galaxy catalogue
\begin{equation}
\hat{\gamma} = \frac{1}{2 S_{\rm{Sh}}} \,\, \epsilon %Krev
\label{eq:gamma_e}
\end{equation}
where the factor $S_{\rm{Sh}} \sim 0.88$ is the average responsivity
of the source galaxies to a shear, see \cite{sheldon01} and references
therein, 
and $\epsilon = (a^2 - b^2) / (a^2 + b^2)$, where $a$ is the semi-major and $b$ the semi-minor axis of the shear galaxy. %Krev
We have redone our main NFW dark matter ellipticity result
using a value $S_{\rm{Sh}}$ which is twenty per cent higher, and find
only a negligible change.  Note that the tangential and cross shears
are invariant with respect to rotation of the cluster coordinates. We
bin the shear galaxy catalogue into square pixels of size $0.4 \times
0.4$ arcmin on the sky, but our main results are insensitive to this
exact value. We average the shear estimates in each pixel.

Fig.~\ref{fig:gammaT} shows the smoothed tangential shear. As
expected, the tangential shear is largest in the cluster center. To
interpret this figure further it is helpful to consider the tangential
and cross shears pattern predicted from popular cluster models,
discussed further in Section~\ref{sec:modelling}. Cluster ellipticity
causes little, if any, cross shear, depending on the cluster profile.
For an elliptical SIE the cross shear is exactly zero. For an
elliptical NFW with a major to minor axis ratio of 0.5 (our best fit
result), a mass of $1\times 10^{14} h^{-1} M_{\odot}$ and cluster
redshift 0.15, the maximum cross shear occurs when approaching the
center of the lens. Just outside our mask radius of $0.5 h^-1$ Mpc,
the maximum cross shear is 0.0012 for a source redshift of 0.3. This
value is 10 per cent of the maximum tangential shear outside the mask,
and is therefore small compared to our uncertainties. %Krev
The main effect
of cluster ellipticity %Krev
is to produce an ellipticity in the tangential shear map. Therefore
the tentative visual indication of some horizontal elongation in this
figure is our first hint of dark matter ellipticity.

To compare the data with models we must incorporate the errors on the
shear estimates. The errors on the shear measurements are given by
\be
\label{eq:uncert}
\sigma^2_{\gamma_{\rm T}} =
\sigma^2_i
  + \sigma^2_{\rm SN}
\ee
where $\sigma_i$ is the uncertainty in the shape measurement due to
the finite number of photons falling in each detector element, plus
detector noise, and $\sigma_{\rm SN}$ is the `shape noise' due to
intrinsic variance in the unlensed galaxy shapes (assumed the same for
all galaxies).

We calculate the shape measurement uncertainty $\sigma_i$ from the
uncertainties in the two components of the ellipticity $e_1$ and $e_2$
using
\be
\sigma_i = \frac{1}{2 S_{\rm{Sh}}} \frac{\sigma_{e1} + \sigma_{e2}}{2}.
\ee
The first factor converts from $e$ to $\gamma$ and the second part
assumes that the uncertainty on the two ellipticity components is
essentially equal and uncorrelated, which is approximately true for
some shear estimators~\citep[e.g. Fig.~2 of][]{bridlekbg02}.  If these
assumptions are true then it follows that the ellipticity uncertainty
on a component is independent of rotation.

We estimate the shape noise $\sigma_{\rm SN}$ by calculating the rms
dispersion in $\hat{\gamma}_T$ values as a function of $\sigma_i$. We
compare this to $\sigma_{\gamma_{\rm T}}$ for various $\sigma_{\rm SN}$
values and find $\sigma_{\rm SN}=0.24$ to be the best fit.

%
%
%\clearpage
\begin{figure*}
%(a)
\includegraphics[width=9cm]{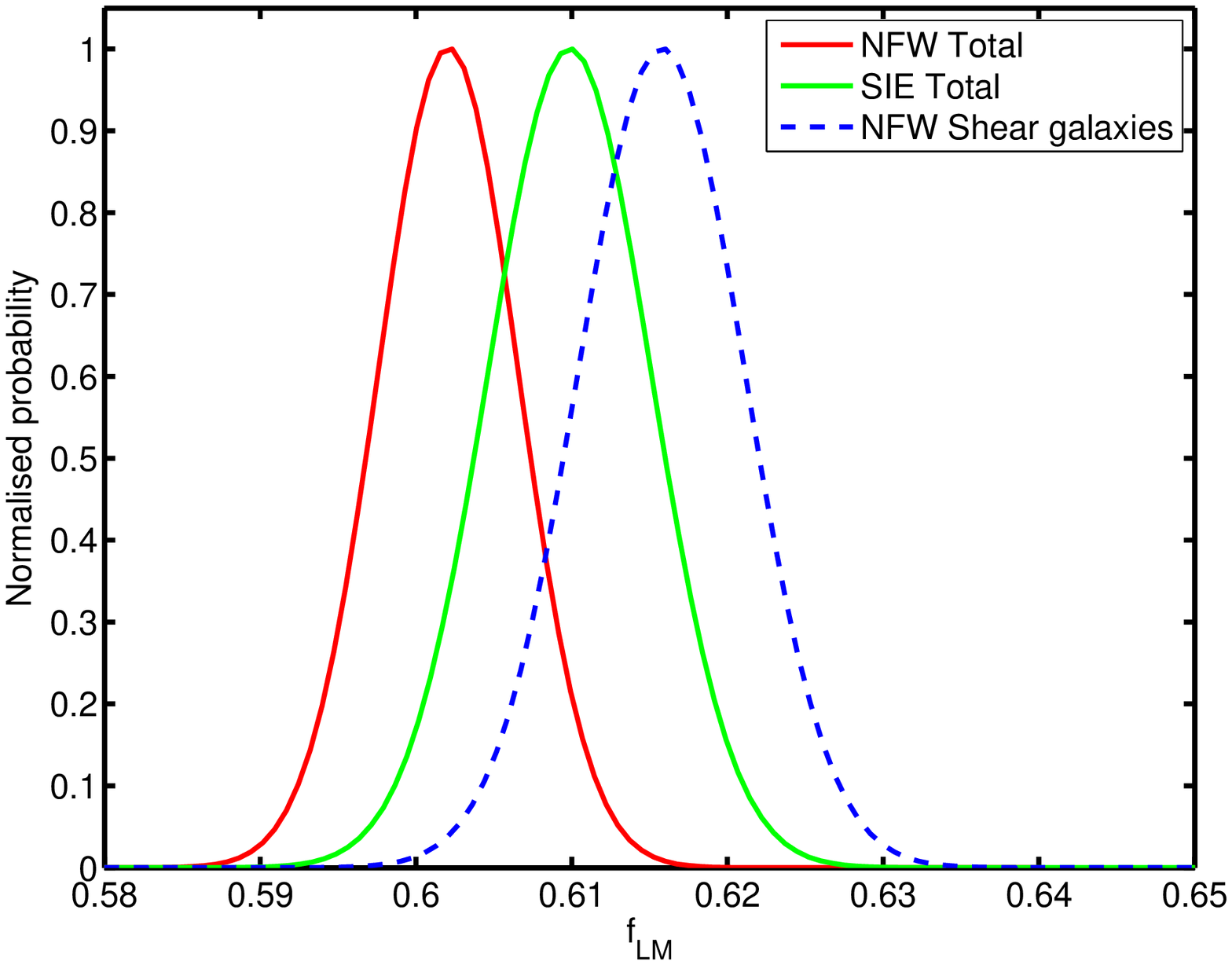}
%(b)
\includegraphics[width=9cm]{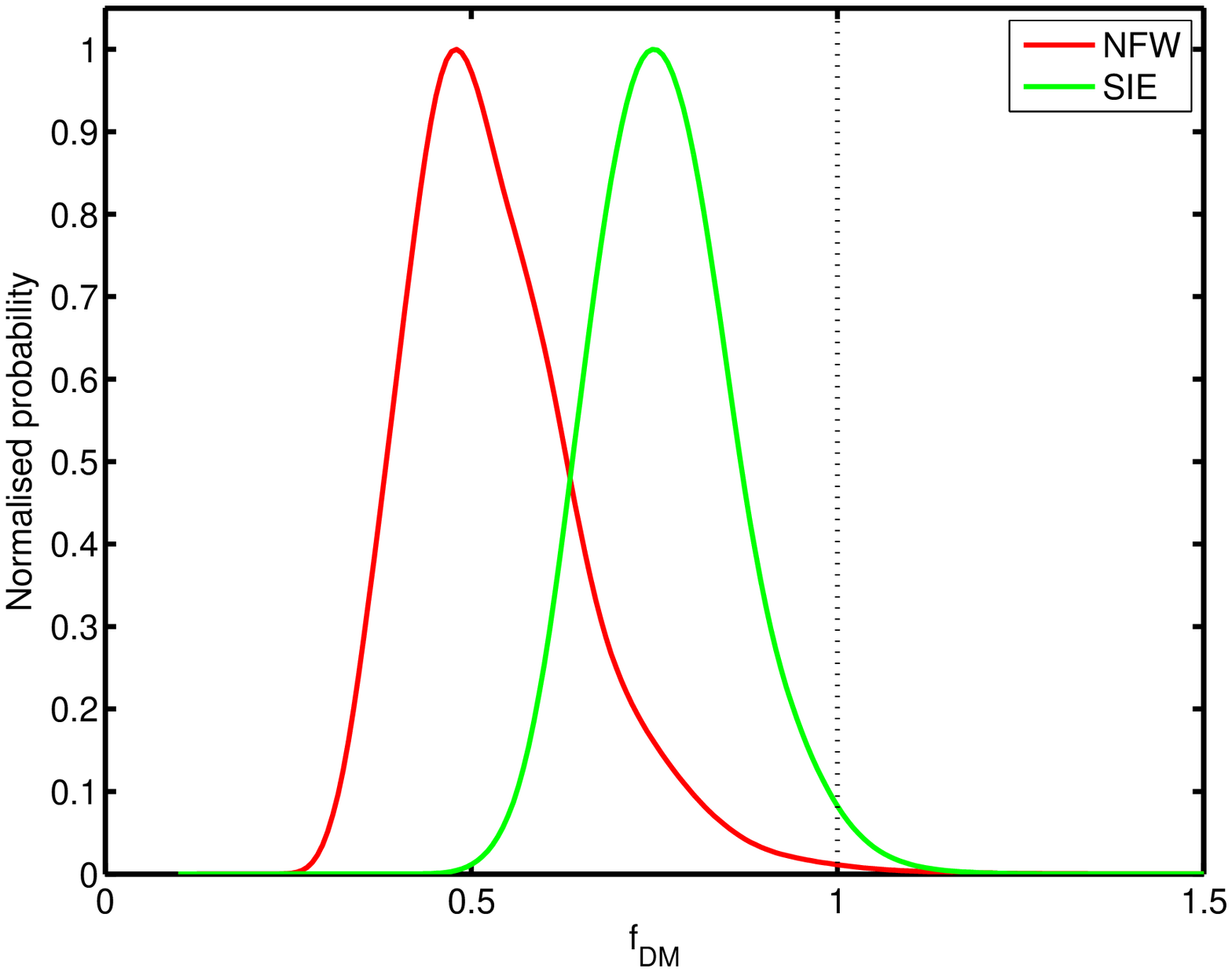}
\caption{ %Panel (a) shows the relative 
  Left panel: Relative probability distribution of the
  axis ratio $f_{\rm LM}=b/a$ for the galaxy number density map from
  the shear catalogue plus member catalogue, thus including all
  cluster members.  This includes clusters at all redshifts for an NFW
  model (dark, red line) and a SIE model (light, green line). The
  dashed line shows the result from shear catalogue galaxies only (for
  an NFW model).  %(b): 
  Right panel: Relative probability distribution for the dark
  matter axis ratio $f_{\rm DM}$.  The dark, red line shows result
  from the NFW model, and the light, green line shows result from the
  SIE model. The dotted vertical line shows $f_{\rm{DM}} = 1$
  corresponding to a circular distribution.  }
  \label{fig:bestfits}
\end{figure*}
%\clearpage
%
%%

To find the error on the shear for each pixel we take the mean of the
errors $\sigma_{\gamma_T}$ and divide by the square root of the number
of galaxies in the pixel. Galaxies will have a range of sizes and
therefore of measurement errors. When we calculate the average shear
in each pixel we do not weight the galaxies according to their
ellipticity errors. This is because this would tend to upweight the
better measured galaxies. As better measured galaxies may
preferentially tend to be cluster members that have leaked into the
shear galaxy catalogue, this might preferentially weight up cluster
members. This would have to be taken into account when removing the
bias on the shear due to cluster member contamination (see
Sec.~\ref{sec:contam}), which would be difficult. We therefore use all
the galaxies in the shear galaxy catalogue without weighting. Due to
the cuts already made in creating the shear galaxy catalogue, the
difference in weight between the noisiest and least noisy galaxies is
only 30\% therefore the weighting would not make a large difference to
our analysis.

\section{Modelling}
\label{sec:modelling}
\subsection{Mass and light distributions}
\label{sec:models}
The shear of galaxies depends on the mass and structure of the cluster
acting as a lens. To model the cluster mass distribution, we use two
alternative theoretical models: a Singular Isothermal Ellipsoid (SIE)
and a Navarro-Frenk-White (NFW) model. The NFW model is preferred from
simulations, but we also include results from the simpler SIE model to
show an extreme and simple example of the dependence of our results on
the cluster profile. %Krev

The SIE model corresponds physically to a distribution of
self-gravitating particles with a Maxwellian velocity distribution
with one-dimensional velocity dispersion $\sigma_v$. The convergence
(normalized mass density) $\kappa = \Sigma/\Sigma_{\rm crit} $ of an
SIE is given by
\be
\kappa = 2 \pi \frac{\sigma_v^2}{c^2} \frac{D_{\rm
    ds}}{D_{\rm s}} r^{-1}
\label{eq:kappa}
\ee
where $r$ is the generalized radius
\be
r = (x^2f + y^2 f^{-1})^{1/2}
\label{eq:genrad}
\ee
and $f = b/a$ is the axis ratio of the ellipse ($b < a$), $\sigma_v$
is the velocity dispersion and $c$ is the speed of light. $x$ and $y$
are coordinates in the plane of the sky, at the cluster redshift. The
distances $D_{\rm ds}$ and ${D_{\rm s}}$ are the angular diameter
distances between the lens (deflector) and the source, and from the
observer to the source, respectively. The SIE peaks sharply in the
central parts, but we mask out the central regions (see
Sec.~\ref{sec:contam}). For the SIE model, we have the simple relation
that the normalized surface density equals the tangential shear
$\gamma_T = \kappa $ \citep{kassiolak93,kormannsb94}.

The NFW model is a more complicated but more realistic model based on
numerical simulations. In order to implement the NFW, we need the
projected mass of the cluster. To calculate the projected mass we use
the equations given in \cite{wrightb00} and \cite{bartelmann96} using
our generalized radius of Eq.~\ref{eq:genrad} to make the cluster mass
distribution elliptical. We use $M_{200}$, the mass enclosed within
the radius at which the density is 200 times the mean density of the
Universe, for consistency with simulations. We derive the
concentration parameter, $c$, where $c\propto M^{\beta}$ according to
Eq.~12 of \cite{seljak00}, where we interpret the virial mass $M$ as
$M_{200}$. 
%Krev We derive the concentration parameter, $c$, as a function of $M_{200}$ using Eq.~12
%Krev of \cite{seljak00} 
We use $\beta=-0.15$, as appropriate for an NFW
model. The shear for an elliptical mass distribution is calculated
using the equations in \cite{keeton01} which are derived from those
in~\cite{schramm90}. A shear map using these equations is illustrated
in Fig.~1 of~\cite{bridlea07}.

We calculate probability as a function of our free parameters in each
redshift bin, and marginalise over all but the axis ratio $f$. We then
obtain a single result for $f$ from combining all the redshift bins by
multiplying the probabilities from different redshift bins together
for each $f$ value. This is the correct calculation if we believe that
the other parameters have different values in each redshift bin, but
that $f$ is the same for all redshift bins. %Krev (inserted as new paragraph?)

\subsubsection{Estimation of the light matter axis ratio ($f_{\rm LM}$) using galaxy positions} %Krev

When stacking the clusters (Sec.~\ref{sec:stackandrot}), we calculated
individual cluster ellipticities based on the cluster galaxies in the
members catalogue. We do not use these results as our measure for the
light matter ellipticity for the stacked cluster because they are
relatively noisy due to the small number of members ($\sim 10$ for the
least rich clusters). Furthermore, we know that the members catalogue
does not in fact contain all the cluster members, and has some
selection criteria that may affect the ellipticity (requirement on
proximity to cluster centre).  We therefore use a $\chi^2$ analysis to
find the light matter ellipticity of the clusters. We model the light
map as coming from (i) non-cluster galaxies which have a constant
density across the postage stamp plus (ii) a contribution from the
cluster galaxies which is assumed to have a galaxy density
proportional to the mass profiles of Eq.~\ref{eq:kappa}
\be
n^{\rm pred}({\bf r}) = K \kappa({\bf r}) + n_0
\label{eq:ngal_pred}
\ee
where $n_0$ is the background level of galaxies per pixel and $K$ is a
constant. Note that we do not assume that the light is some constant
multiple of the mass, only that the light map is proportional to an
SIE or NFW profile (which may have different parameters than the dark
matter distribution). We do not tie the dark and light map parameters
together because we wish to investigate whether the dark and light
distributions both have the same ellipticity. 
%For practical purposes Krev
%we fix the mass at $M_{200}=10^{15} h^{-1} M_{\odot}$ in this
%calculation, but the exact value has very little effect since the
%mass-to-light ratio is a free parameter (there is a complete
%degeneracy for the SIE model). 
For practical purposes we fix the mass at 
$M_{200}=10^{14} h^{-1} M_{\odot}$
%$M_{200} = 10^{14} \ h^{−1} M_{\odot}$ 
in this calculation, which corresponds approximately to
clusters of the mean richness we used \citep[][ Table 6]{johnston}. The
value used affects the concentration parameter and therefore the mass
profile of the cluster, which affects the weighting of the map. If we
use a value a factor of ten higher our light ellipticity results
change by less than one sigma and in any case our main results, on the
dark matter ellipticity, are changed imperceptibly because the
uncertainties on those are dominated by the much larger uncertainty on
the dark matter quantities.  %Krev

We calculate probabilities in the resulting three dimensional space %Krev
($f_{\rm LM}$, $K$ and $n_0$) by calculating a $\chi^2$ between the predicted
number (Eq.~\ref{eq:ngal_pred}) and the observed number
%Krev
\be
\chi^2 = \sum_i \frac{(n_{\rm pred}({\bf r}_i) - n_{\rm obs}({\bf r}_i))^2}{\sigma_{n_i}^2}
\ee
and calculating probabilities from this;
${\rm Pr} = e^{\chi^2/2}$.%Krev

Our assumption is that the errors are Poisson, therefore %Krev
 in the limit of large numbers the error on the number of galaxies, as
 used in the $\chi^2$ calculation, is
$\sigma_{n_i} = \sqrt{n^{\rm pred}({\bf r}_i)}$.%Krev 
We calculate our main results for each of the
three stacked light maps shown in Fig.~\ref{fig:numbgals}.

\subsubsection{Estimation of the dark matter axis ratio ($f_{\rm DM}$) from the stacked shear map} %Krev

To estimate the ellipticity of the dark matter distribution we
calculate probability as a function of cluster axis ratio $f_{\rm DM}$ and
cluster mass. We calculate the probability from the $\chi^2$ between
the observed and predicted shears, using the uncertainty on the shear
values from Eq.~\ref{eq:uncert}. We marginalize over the cluster mass
with a flat prior to obtain the probability as a function of the 2D
axis ratio $f_{\rm DM}$.%Krev 

\subsection{Redshift distributions}
As discussed in Sec.~\ref{sec:data}, we divide our cluster sample up
into four redshift bins. Due to the large photometric redshift
uncertainties we decided not to use the redshift information in the
shear galaxy catalogue.  Therefore the `shear galaxies' may be in
front of, behind, or part of the cluster. The shear for each
lens-source pair depends on the redshift of both the lens (cluster)
and the source (shear galaxy). To calculate our theoretical model, we
need a prediction for the distance ratio in Eq.~\ref{eq:kappa} at each
possible redshift.  This must be averaged, weighted by the number of
galaxies at each redshift. In other words, we need to calculate
\begin{equation}
\Big\langle 
\frac{{D_{\rm{ds}}}}
{{D_{\rm{s}}}}
%Krev
 \Big\rangle =
\frac{\int_{z_L}^{\infty} (D_{\rm{ds}}/ D_{\rm{s}}) \ n(z_s) \ {\rm{d}}z_s
}
{\int_0^{\infty}  \ n(z_s) \ {\rm{d}}z_s}.
\label{eq:dist_av}
\end{equation}
Note that the integration in the nominator starts at the lens
redshift, $z_L$, so that galaxies between us and the lens do not
contribute to the shear signal, as they are not influenced by the
presence of the cluster. We estimate $n(z_s)$ from Fig.~3 and Eq.~8 in
\cite{sheldon01}, and as a result we use $z_c = 0.22$.  However our
results are quite insensitive to these numbers because we focus only
on the ellipticity of the dark matter halo and not on its mass.

We obtain $\langle D_{\rm{ds}}/D_{\rm{s}} \rangle= [0.51, 0.37, 0.26,
0.17]$ for the four redshift bins.  The values are low for high
redshift bins because a large fraction of the galaxies are between us
and the cluster, and therefore do not contribute to the shearing
effect. The calculation of $D_{\rm{ds}}/D_{\rm{s}}$ is approximate
because we assume all clusters to be located at the center of their
redshift bin. However, this does not affect the ellipticity of the
theory prediction. The distance ratio is incorporated into the
predictions using Eq.~\ref{eq:kappa}.

\subsection{Cluster decontamination}
\label{sec:contam}
Because spectroscopic redshifts are not available for all shear catalogue
galaxies, there will always be a degree of contamination by cluster
members in the shear signal (as seen in Fig.~\ref{fig:numbgals}(a)).
Since we assume that the cluster members have no systematic alignment
(but see Sec.~\ref{sec:conclusions} for an assessment of the
implications of this assumption), %Krev
members that have leaked into the shear galaxy catalogue will tend to
dilute the shear signal. We correct for this dilution in our
analysis. The corrected shear is given by
\be
\gamma_{\rm{cor}} = \frac{n({\bf r})}{n_0} \hat{\gamma}
\label{eq:contam}
\ee
where $n({\bf r})$ is the total number of galaxies in the shear galaxy
catalogue a two dimensional position $\bf{r}$ from the center, $n_0$
is the number of galaxies not in the cluster (see
Eq.~\ref{eq:ngal_pred}) and $\hat{\gamma}$ is the observed,
uncorrected shear. We use the best fit $n_0$ value from the $\chi^2$
fit to the light matter distribution. Inside the cluster, we have
$n({\bf r}) > n_0$, so the observed shear will be boosted by correcting
for the contamination.

We mask the central regions from our analysis for several reasons.
(i)
In the very central regions, cluster members will obscure the
shear galaxies.
(ii)
The cluster center may be incorrect and thus the central parts may appear erroneously circular after the stacking.
(iii)
There is actually an uncertainty associated with the correction
factor that we have not taken into account. This error is due to
Poisson fluctuations in the true number of non-cluster members per
pixel, and will be more significant when the value of the correction
factor is large, which occurs in the central region where the observed
number of galaxies in the shear galaxy catalogue peaks sharply.
For all these reasons, we mask out the central region using a circular
mask with $r_{\rm{mask}} = 0.5 h^{-1}$ Mpc. The radius of the mask was
set where the correction factor increases above 1.5 (as calculated for
$0.20 < z \le 0.25$), shown by the
dashed
circle in
Fig.~\ref{fig:numbgals}. This corresponds to 3.3 arcminutes at
$z=0.225$.

%
%\clearpage
\begin{table*}
\begin{center}
\caption{Axis ratio results for the light maps}
\label{tab:lmresults}
\begin{tabular}{|c|c|c|c|}
\hline
Population & $z$ & SIE ($f_{\rm LM}$) & NFW ($f_{\rm LM}$)\\
\hline
Shear galaxies &$0.10 \le z \le 0.15$ & 0.548 $+$ 0.013 $-$ 0.013& 0.620 $+$ 0.015
$-$ 0.015\\
Total &$0.10 \le z \le 0.15$ & 0.555 $+$ 0.014 $-$ 0.014& 0.614 $+$ 0.012 $-$ 0.013\\
Members &$0.10 \le z \le 0.15$ & 0.302 $+$ 0.006 $-$ 0.006& 0.544 $+$ 0.006 $-$
0.007\\
\hline
Shear galaxies &$0.15 \le z \le 0.20$ & 0.672 $+$ 0.011 $-$ 0.011& 0.649 $+$ 0.012
$-$ 0.012\\
Total &$0.15 \le z \le 0.20$ & 0.657 $+$ 0.012 $-$ 0.011& 0.631 $+$ 0.010 $-$ 0.010\\
Members &$0.15 \le z \le 0.20$ & 0.261 $+$ 0.005 $-$ 0.005& 0.475 $+$ 0.006 $-$ 0.006\\
\hline
Shear galaxies &$0.20 \le z \le 0.25$ & 0.581 $+$ 0.012 $-$ 0.012& 0.583 $+$ 0.011
$-$ 0.010\\
Total &$0.20 \le z \le 0.25$ & 0.578 $+$ 0.011 $-$ 0.010& 0.575 $+$ 0.009 $-$ 0.009\\
Members &$0.20 \le z \le 0.25$ & 0.254 $+$ 0.004 $-$ 0.004& 0.456 $+$ 0.005 $-$ 0.005\\
\hline
Shear galaxies &$0.25 \le z \le 0.30$ & 0.637 $+$ 0.010 $-$ 0.009& 0.619 $+$ 0.008
$-$ 0.008\\
Total &$0.25 \le z \le 0.30$ & 0.624 $+$ 0.008 $-$ 0.008& 0.602 $+$ 0.007 $-$ 0.007\\
Members &$0.25 \le z \le 0.30$ & 0.235 $+$ 0.003 $-$ 0.003& 0.434 $+$ 0.004 $-$ 0.004\\
\hline
\hline
Joint (total)  & $0.10 \le z \le 0.30$ & 0.610 $+$ 0.005 $-$ 0.005& 0.602 $+$ 0.004 $-$ 0.005\\
\hline
\end{tabular}
\end{center}
\end{table*}
\bigskip
\begin{table}
\caption{Axis ratio results for the dark matter distribution.
}
\label{tab:dmresults}
\center{
\begin{tabular}{|c|c|c|}
\hline
$z$ &SIE ($f_{\rm DM}$) &NFW ($f_{\rm DM}$)\\
\hline
$0.10 \le z \le 0.15$ & 0.754 $+$ 0.230 $-$ 0.186& 0.614 $+$ 0.400 $-$ 0.176\\
$0.15 \le z \le 0.20$ & 0.522 $+$ 0.159 $-$ 0.115& 0.269 $+$ 0.158 $-$ 0.054\\
$0.20 \le z \le 0.25$ & 0.958 $+$ 0.251 $-$ 0.199& 0.599 $+$ 0.379 $-$ 0.208\\
$0.25 \le z \le 0.30$ & 0.853 $+$ 0.306 $-$ 0.215& 0.614 $+$ 0.444 $-$ 0.266\\
\hline
\hline
Joint analysis & 0.747 $+$ 0.102 $-$ 0.094& 0.480 $+$ 0.136 $-$ 0.086\\ 
\hline
\end{tabular}
}
\bigskip
\end{table}
%\clearpage
%
%

\section{Results}
\label{sec:results}
We now present our results on the stacked cluster ellipticity,
focussing first on the NFW profile. %Fig.~\ref{fig:bestfits}(a) 

\subsection{Results for the light matter axis ratio ($f_{\rm LM}$)} %Krev
The left panel of Fig.~\ref{fig:bestfits} shows the one-dimensional
relative probability of the axis ratio $f_{\rm LM}$, marginalised over $K$ and
$n_0$ (see Eq.~\ref{eq:ngal_pred}). The solid lines represent results
from using the total cluster members, i.e.~galaxies in both the shear
galaxy catalogue and cluster member catalogue (corresponding to the
right hand panel of Fig~\ref{fig:numbgals}). The blue, dashed line
shows results from using only the (members-removed) shear catalogue
galaxies, corresponding to the left hand panel of
Fig.~\ref{fig:numbgals}, for the NFW model. This shows that the light
matter distribution is clearly elliptical with an axis ratio of $f_{\rm LM}
\sim 0.6$. The errors on $f_{\rm LM}$ are calculated by finding the 68\%
iso-probability limits from the probability distribution $P(f_{\rm LM})$ after
marginalizing over the other fit parameters.  We find an error of
$\sim 0.005$.

Table~\ref{tab:lmresults} shows three light matter ellipticity results
for each redshift bin and for each theoretical model (SIE and NFW):
(i) for the shear catalogue galaxies with cluster members removed,
(ii) for the total cluster members (iii) for the cluster members and
(members from the cluster catalogue plus the extra members included in
the shear galaxy catalogue).  The results from the NFW and SIE are
qualitatively similar.
%Krev 
%We calculate probability as a function of our free parameters in each
%redshift bin, and marginalise over all but the axis ratio $f$. We then
%obtain a single result for $f$ from combining all the redshift bins by
%multiplying the probabilities from different redshift bins together
%for each $f$ value. This is the correct calculation if we believe that
%the other parameters have different values in each redshift bin, but
%that $f$ is the same for all redshift bins.
% (merge these two paragraphs?)
The last line in Table \ref{tab:lmresults} shows the joint results
combining all bins, for the total cluster members.

There is a clear detection of ellipticity based on the number density
of shear catalogue galaxies alone. We see that the distribution of
galaxies in the members catalogue is more elliptical than that of the
shear catalogue. This is not surprising because we stacked the
galaxies according to the members catalogue. Even if the galaxies in
the members catalogue had been sampled from a circular distribution
then the finite number of galaxies would provide a rotation direction
so we would have in effect stacked the random noise to produce an
ellipticity for the cluster members, while making the light map from
the shear catalogue more circular.

This extra induced ellipticity could be simulated, but in fact there
is no need because the axis ratio from the shear galaxy catalogue
alone is very similar to that based on the total catalogue
containing both the shear catalogue \emph{and} the cluster member
catalogue. Therefore we do not consider the results from only the
cluster member catalogue any further.

The similarity of the ellipticities from the shear galaxy catalogue
and the total catalogue is largely due to the fact that there are very
few galaxies from the members catalogue outside our mask radius. The
important conclusion for our paper is that any misalignment in the
rotating and stacking will affect the light and dark matter
ellipticity the same.  The ellipticity observed in the light map makes
us believe that the rotation angles are not completely random and
therefore that we can hope to detect some ellipticity in the dark
matter, if indeed the dark matter halo is elliptical.

\subsection{Results for the dark matter axis ratio ($f_{\rm DM}$)}%Krev
Ellipticity results for the dark matter analysis can be found in Table
\ref{tab:dmresults}, for each individual redshift bin as well as for
the joint result combining all redshift bins. For the NFW model, the
joint result of $f_{\rm DM} = 0.48^{+0.14}_{-0.09}$ excludes a circular mass
distribution ($f_{\rm DM} =1$) by over $3 \sigma$ (if the probability
distribution is Gaussian).  The lowest axis ratio is in the second
redshift bin ($0.15 < z < 0.20$). We re-analyzed redshift bin 2 by
dividing it into sub-bins, but found the result unchanged on removing
the clusters with the lowest axis ratio. The joint result for the
remaining redshift bins was then $f_{\rm DM} = 0.607^{+ 0.21}_{- 0.14}$. This
is a weaker detection than our final result including the second
redshift bin, but still a tentative detection of dark matter
ellipticity.

%Fig.~\ref{fig:bestfits}(b) 
The right panel of Fig.~\ref{fig:bestfits} shows the probability as a
function of axis ratio for the dark matter distribution. The dark
(red) line shows the result from using an NFW model, and the light
(green) line shows the result from an SIE model. The vertical dotted
line is showing $f_{{\rm DM}} = 1$, which represents a circular mass
distribution.

We can see from the figure that the probability is not totally
Gaussian, with a tail to larger $f_{\rm DM}$ values. (Note that an axis ratio
greater than unity corresponds to an ellipse aligned along the $y$
axis.) We find that 99.6 per cent of the probability is below $f_{\rm DM}=1$;
therefore we consider this to be a reasonable detection of dark matter
ellipticity. Note that the best fit NFW result is quite a bit more
elliptical than the SIE result, but the NFW result has a longer tail
extending to higher $f_{\rm DM}$-values than the SIE. For the SIE model 98.2
per cent of the probability is below $f_{\rm DM}=1$. The NFW is a more
realistic profile, so we trust these results most.

%
%\clearpage
\begin{figure}
\includegraphics[width=8cm]{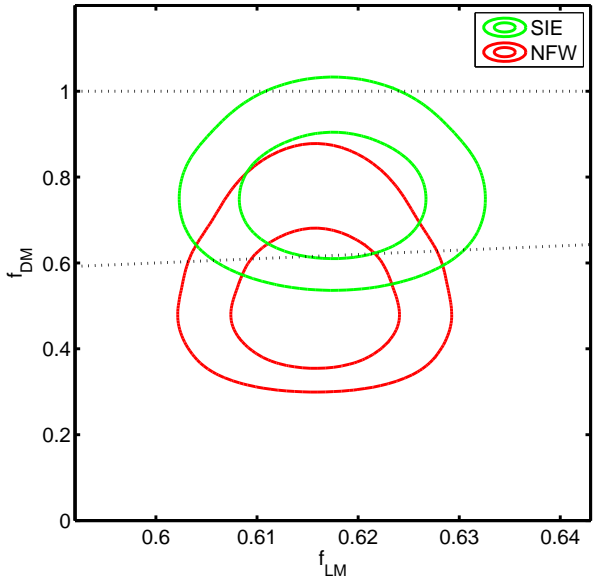}
\caption{ 
  68\% and 95\% contours of the two-dimensional probability distribution for
  the axis ratio of the dark matter ($f_{{\rm DM}}$)versus the axis
  ratio of the total galaxy number density distribution ($f_{{\rm
      LM}}$, representing the light matter).  Light (green) contours
  show the result from using an SIE profile as modelling the cluster,
  and dark (red) contours show result from using an NFW profile. The
  horizontal dotted line shows $f_{{\rm DM}} = 1$, and the dotted line
  rising towards the right shows $f_{{\rm DM}} = f_{{\rm LM}}$. }
  \label{fig:res_cont}
\end{figure}

\begin{figure}
\includegraphics[width=8cm]{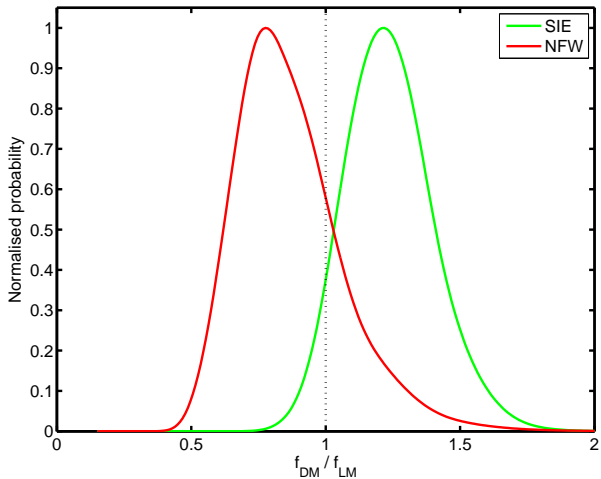}
  \caption{
    One-dimensional probability distributions as a function of the
    ratio $f_{{\rm DM}}/f_{{\rm LM}}$. Light (green) line is for the
    SIE result while dark (red) line is for the NFW result. }
  \label{fig:res_dmlm}
\end{figure}
%\clearpage

Fig.~\ref{fig:res_cont} shows 68\% and 95\% contours of the
two-dimensional probability distribution for the axis ratio of the
dark matter ($f_{{\rm DM}}$) versus the axis ratio of the light matter
($f_{{\rm LM}})$. The dotted lines also shown on the plot are $f_{{\rm
    DM}} = 1$ (horizontal dotted line) and $f_{{\rm DM}} = f_{{\rm
    LM}}$ (dotted line rising towards the right). The two colors shown
represent the different profiles used as cluster models: light (green)
contours are for the SIE model, and dark (red) contours are for the
NFW. We see that $f_{{\rm DM}} = 1$, that is, a circular dark matter
distribution, crosses just within the outermost (95\%) contour for the
SIE model and the NFW contours are both below the $f_{{\rm DM}} = 1$
line.

Fig.~\ref{fig:res_dmlm} shows the one-dimensional probability
distribution as a function of the ratio of axis ratios of the dark and
light matter: $f_{\rm DM}/ f_{\rm LM}$. Again, the light (green) line
is for the SIE model profile and the dark (red) line is for the NFW
model. The vertical dotted line shows where we have $f_{\rm DM} =
f_{\rm LM}$, that is; the axis ratio as deduced from the shear values
equals the axis ratio as deduced from the distribution of the cluster
galaxies. Both profiles are consistent with having the same
ellipticity in the light and in the dark matter, $f_{\rm DM}/ f_{\rm
  LM}=1$. We find $ f_{\rm DM}/ f_{\rm LM} = 0.91^{+ 0.19}_{- 0.25}$
for the NFW model and $f_{\rm DM}/ f_{\rm LM} = 1.26^{+ 0.16}_{- 0.17}$
for the SIE. The higher SIE result is driven by the larger $f_{\rm
  DM}$ for the SIE, which makes the dark matter distribution appear
more round.

\subsection{Dependence on close neighbor distance}
%
%\clearpage
\begin{table*}
\caption{Effect of neighbor removal on measured ellipticity.
Results are shown for the NFW profile only.
}
\center{
\begin{tabular}{|c|c|c|c|}
\hline
Cut/ $h^{-1}$ Mpc  & No. clusters & Light matter $f_{\rm LM}$ & Dark matter $f_{\rm DM}$\\
\hline
2.5 & 6934 & 0.587 $+$ 0.004 $-$ 0.004 & 0.459 $+$ 0.155 $-$ 0.077\\
5   & 4281 & 0.602 $+$ 0.004 $-$ 0.005 & 0.480 $+$ 0.136 $-$ 0.086\\
7.5 & 2542 & 0.625 $+$ 0.006 $-$ 0.006 & 0.614 $+$ 0.205 $-$ 0.139\\
\hline
\end{tabular}}
%\tablenotetext{}{
\label{tab:neighb}
\end{table*}
%\clearpage

%
We made a decision to exclude all clusters with a more massive
neighbor within an angular size corresponding to $5 h^{-1}$ Mpc at the
cluster redshift. In this section we test how dependent our results
are on this decision. Table~\ref{tab:neighb} shows that our main
result changes little as the cut is made 50 per cent smaller or
larger. As expected, when a larger radius is used, fewer clusters
survive in the catalogue, therefore the uncertainties become larger.
Using a smaller radius does increase the ellipticity which is to be
expected if the main effect is to include more physically associated
clusters which are more likely to lie along the major axis of the
cluster, perhaps due to formation along an intervening filament. We
find similar results for the light ellipticity. Our main results are
unchanged by changing the cluster isolation criterion.

\subsection{Misalignment simulations}
\label{sec:mis}
In order to interpret our results we need to take into account
possible misalignments during stacking.
When we rotate clusters according to the distribution of the cluster
members before stacking, we assume that the orientation of the light is
correlated with the orientation of the mass. A result of $f_{\rm DM}$
consistent with 1 could indicate \emph{either} a circular mass
distribution \emph{or} a random alignment between the light used for
stacking and the dark matter. In the latter case, the circular result
would be caused by the stacking of many clusters with different
misalignments between the dark and the light matter.

In order to quantify this effect we simulate many clusters all with
the same input axis ratio $f_{\rm{in}}$. The misalignment angles
between the light and the dark matter for the simulated clusters are
random, with a standard deviation $\sigma_{\theta}$. For each value of
$f_{\rm{in}}$ and $\sigma_{\theta}$, we rotate and stack the clusters
in the same way as for the SDSS data. Therefore if $\sigma_{\theta}$
is large, the input ellipticity will be smeared out a great deal and
output axis ratio $f_{\rm out}$ will be close to unity. If
$\sigma_{\theta}$ is small the output ellipticity will be more similar
(or in the case of zero misalignment, equal) to $f_{\rm{in}}$.

\begin{figure}
\scalebox{1.1}{\includegraphics[width=8cm]{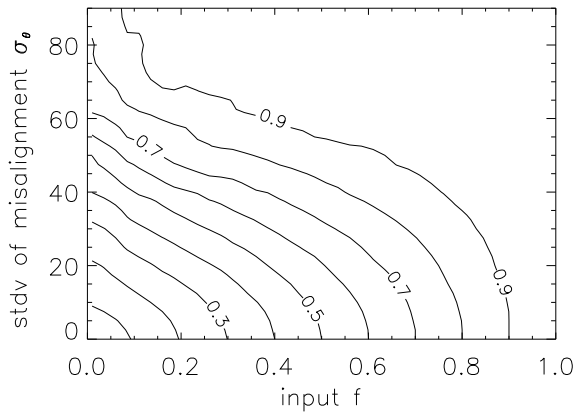}}
\caption{ 
  Contours show output axis ratios for simulated misalignments between
  the light and dark matter. Axes show input axis ratio $f_{\rm{in}}$
  and standard deviation $\sigma_{\theta}$ (in degrees) of the angle
  distribution of the simulated clusters.  }
  \label{fig:misal}
\end{figure}

Fig.~\ref{fig:misal} shows the output dark matter axis ratio as a
function of degree of misalignment $\sigma_{\theta}$ and input axis
ratio $f_{\rm{in}}$. In order for the input ellipticity to be very
elliptical ($f \rightarrow 0$), the misalignment between the dark and
light matter must be $\sim 50^\circ$ for our dark matter $f_{\rm DM}$ of $\sim
0.5$ (see Table \ref{tab:dmresults}).  We conclude that the
misalignment angle must be less than $\sim 50^\circ$. If there is any
misalignment between the light and the dark matter, the dark matter
will be even more elliptical than the results shown in the table and
can be read off Fig.~\ref{fig:misal} for a given misalignment angle.

%
%
%\clearpage
%\clearpage
%

\section{Conclusions and discussion}
\label{sec:conclusions}
We have used galaxy clusters from the catalogue of \cite{bencat} and
shear maps as used in \cite{sheldon04} to investigate the light and
dark matter ellipticities of galaxy cluster halos.  We rotate the
selected clusters so that their major axes are aligned, and stack them
according to the method which is described in \cite{nat2000} for
galaxy-galaxy lensing. This is the first time that this method has
been applied to cluster lensing.

Through the pattern imprinted on the shear maps by intervening massive
cluster halos, we have detected a dark matter ellipticity of these
halos at a 99.6\% level, with an axis ratio of $f_{\rm DM} =
0.48^{+0.14}_{-0.09}$ from a joint $\chi^2$-analysis for an NFW model,
using 4281 clusters between $0.10 < z \le 0.30$. We have corrected for
dilution of the shear signal by cluster members left in the shear
galaxy catalogues but have masked out the central areas where this
correction factor grows too large.

The light matter distribution of the clusters, as traced by the number
density of individual cluster members, is also clearly elliptical,
with a joint axis ratio $f_{\rm LM} = 0.60^{+0.004}_{-0.005}$ for the NFW
model.  Using the shear catalogue alone gives very similar results to
using the shear catalogue concatenated with the cluster member
catalogue, which means that we are essentially comparing like with
like when comparing dark and light map ellipticities. This is because
both the light matter (shear catalogue) and the dark matter have been
stacked in the same way.

We find that the ellipticity of the dark matter distribution is
consistent with the ellipticity of the galaxy number density
distribution. Our result is limited by the uncertainty in the
ellipticity of the dark matter distribution. The results for the NFW
and SIE agree within the errors, but any differences could be
attributed to a changing ellipticity with radius.  We have not
attempted to measure this effect since the uncertainties are too
large.

Shear maps from neighboring clusters will influence each other,
making the pattern more elliptical or less, depending on the redshift
of the neighbors. In order to reduce this effect as much as possible,
we use only the cluster with the highest number of members
($N_{\rm{gals}}^{r200}$) when two (or more) clusters are closer
together than $5 h^{-1}$ Mpc, as seen on the sky. We find that
increasing or decreasing the minimum distance to a close neighbor by
50 \% does not significantly affect our main results.

We have also simulated the effect that a possible misalignment between
the light and dark matter could have, and concluded that, according to
our results, the light and dark matter must be misaligned by less than
$\sim 50^\circ$.

Our measurement is very insensitive to overall calibration biases in
shear measurements, since these are degenerate with cluster mass,
rather than cluster ellipticity. It is unlikely that biases in shear
measurement would vary with angle around the cluster: residual point
spread function anisotropies would be oriented at random with respect
to the cluster major axis and would cancel out on stacking; the
cluster member light may leak into the postage stamps used to measure
shears of background galaxies, but we remove the central region where
the number of confirmed cluster members is significant.

Possibly the biggest potential systematic is intrinsic alignment of
cluster members pointing at the cluster center
\citep{ciottid94,kuhlendm07,pereirabg08,knebe08,faltenbacherea08}
which would be a problem for us because we have not made a significant
attempt to remove cluster members from our analysis and our background
catalogue is not particularly deep. A thorough assessment of this
effect is beyond the scope of the current work.  To first order this
would make our observed shear maps less elliptical since the
contamination by cluster members is greatest along the cluster major
axis, and thus the strong gravitational shears expected along this
axis will be partially cancelled out by the cluster members which have
the opposite ellipticity since they tend to point at the cluster
center.  However a full assessment of this effect would also have to
take into account the variation in intrinsic alignment with respect to
the cluster major axis, which appears to be more complicated
\citep{kuhlendm07} and possibly weakens the degree of cancellation
preferentially along the cluster major axis.

In \cite{hopkins05}, the authors use a large-scale, high-resolution
$N$-body simulation to predict cluster ellipticities and alignments in
a $\Lambda$CDM universe. They find an ellipticity
\be
\langle \epsilon \rangle = 1 - \langle f \rangle = 0.33 + 0.05z.
\label{eq:hopkins}
\ee
for the redshift range $0<z<3$. This redshift evolution is negligible
for the redshift range considered here, and due to the large
uncertainties we do not try to detect any trends with redshift. For a
redshift in the middle of the redshift range of our cluster sample
($z=0.2$), this formula yields an axis ratio of $f = 0.66$ which is in
good agreement with our result of $f_{\rm DM}=0.48^{+0.14}_{-0.09}$.

\cite{ho06} have used numerical simulations of cluster formation with
the aim of investigating the possibility of using cluster
ellipticities as a cosmological probe. They find that the mean
ellipticity of high mass clusters is approximated by
%
%Krev\be
%Krev\bar{e}(z=0) = 0.245 - 0.070 \ \sigma_{8,0} + 0.020 \ \Omega_{m,0}.
%Krev\ee
%
\be
\bar{e}(z) = 0.245 \ \Bigl[ 1 - 0.256 \ \frac{\sigma_{8}(z)}{0.9} + 0.00246 \ \frac{\Omega_m}{0.3} \Bigr].
\ee
Using $\sigma_{8}(z) = 0.8$ and $\Omega_{m}=0.3$, this gives an
ellipticity of $e \sim 0.2$, or $f \sim 0.8$. This result is more
circular than our main result. However, to make a proper comparison of
our result with theory it would be necessary to make a theoretical
prediction that takes into account our observation method, especially
the overlaps in the shear field due to close neighbors and the impact
of selecting the most isolated clusters.

\cite{hoekstrayg04} report on a first weak-lensing detection of the
flattening of galaxy dark matter halos, using data from the
Red-Sequence Cluster Survey. They find an average galaxy dark matter
halo ellipticity of $\langle \epsilon \rangle = 0.33^{+0.07}_{-0.09}$.
They also find a detection that dark matter halos are rounder than the
light map. In a later work, \cite{mandelbaumea06} did not find a
definite detection of this effect in the larger SDSS dataset. However,
\cite{parker07}, measuring the ratio of tangential shear in regions
close to the semi-minor versus that close to the semi-major axes of
the lens, find some evidence of a halo ellipticity of $\sim 0.3$ using
early data from the Canada-France-Hawaii Telescope Legacy Survey
(CFHTLS). %Krev
\cite{mandelbaumea06} commented that the stronger signal in clusters
of galaxies means that there is more chance of making a detection of
ellipticity in this higher mass data set. %Krev
Therefore we have confirmed the detection of dark matter halo
ellipticity, extending the measurement to cluster scales. We find no
evidence for different ellipticities for the light and dark matter
distribution on cluster scales.

\cite{mandelbaumea06} focus on measuring the ratio of (dark matter)
halo ellipticity to galaxy (light) ellipticity $\epsilon_h/\epsilon_g$
where $\epsilon =(a^2 - b^2)/(a^2 + b^2)$. They find
$\epsilon_h/\epsilon_g=0.60 \pm 0.38$ (68 per cent confidence) for
elliptical galaxies, which is most comparable with our result of
$f_{\rm DM}/f_{\rm LM} = 0.91^{+0.19}_{-0.25}$ (68 per cent
confidence) where $f=b/a$. Repeating our calculation using the
different ellipticity parameters we find $\epsilon_h/\epsilon_g =
1.37^{+0.35}_{-0.26}$. It is expected that our value is greater than
unity, since we find that the dark matter is more elliptical (lower
axis ratio) than the light matter. However it is not significantly
larger. The result of \cite{mandelbaumea06} is the opposite side of
unity, but the difference is not significant and we note that our
result is for clusters and that of \cite{mandelbaumea06} is for
galaxies.

The anisotropy of the lensing signal around individual galaxies, if
definitely detected, has been said to pose a serious problem for
alternative theories of gravity \citep{hoekstrayg04}.  For galaxy
clusters it is more complicated, as the dominant source of baryons is
the intracluster gas. Lensing by clusters has been found to pose a
problem for Modified Newtonian Dynamics (MOND), as there seems to be a
need to include a dark matter component \citep[]{sanders03,
  takahashi07}.

Because the gas - the dominant baryonic component of clusters - is
collisional, we can suppose that on cluster scales it will be less
elliptical than the light. We may end up concluding that either (i)
the gas distribution is elliptical (which we would not expect) or (ii)
our result is inconsistent with MOND. However, further study and
simulations of this is clearly needed.

\acknowledgements 
We would like to thank Erin Sheldon and Benjamin
Koester for providing us with galaxy catalogues and for very helpful
comments and suggestions. We would also like to thank the anonymous referee for many
helpful comments and suggestions. %Krev
We would also like to thank 
Timothy McKay,
Jochen Weller, H\aa kon Dahle, Andreas Jaunsen,
Margrethe Wold,
Shirley Ho,
Eduardo Cypriano, Laurie Shaw, Richard Cook, David Sutton,
Andrey Kravtsov,
\O ystein Elgar\o y,
Morad Amarzguioui,
Terje Fredvik
and Stein Vidar Hagfors Haugan for helpful
discussions.
AKDE acknowledges support from the Research Council of Norway, Project
No.~162830. SLB acknowledges support from the Royal Society in the
form of a University Research Fellowship.

%%%%%%%%%%%%%%%%%% Bibliography %%%%%%%%%%%%%%%%%%%%%%%%%%%%%%%%%%%

%\bibliographystyle{plain}
%\bibliography{dme_paper_lastv}

\end{document}